# Observations of Random Walk of the Ground in Space and Time

*Vladimir Shiltsev*

*Fermi National Accelerator Laboratory, PO Box 500, Batavia, IL 60510, USA*

*Abstract*

We present results of micron-resolution measurements of the ground motions in large particle accelerators over the range of spatial scales $L$ from several meters to tens of km and time intervals $T$ from minutes to several years and show that in addition to systematic changes due to tides or slow drifts, there is a stochastic component which has a "random-walk" character both in time and in space. The measured mean square of the relative displacement of ground elements scales as $dY^2 \approx ATL$ over broad range of the intervals, and the site dependent constant $A$ is of the order of $10^{-5\pm1}$ μm$^2$/(s·m).

Ground motion is often characterized by a combination of three components – systematic trends due to long-term geological motions, one or more periodic components, such as Earth tides, daily and seasonal changes associated with temperature or air pressure variations, and stochastic movements [1]. The stochastic component is usually less correlated in space, less persistent in time and less predictable than the first two while not necessarily smaller in amplitude and, thus, it often poses the biggest concern. Fractal properties of the stochastic component of the ground motion have long been known to geophysicists, (see, e.g., [2]). For example, the topography analysis shows that the variance of the difference of elevations of two points separated by distance $L$ scales as $dY^2 \propto L^{\gamma}$, $\gamma \approx 1$ [3]. Similarly, the variance of the relative motion of two points over a time interval $T$ often follows the power-law $dY^2 \propto T^{\alpha}$, where $\alpha \approx 1$ [4]. To explore temporal and spatial properties of ground motion simultaneously, studies of dynamics at numerous points are needed. That is where high precision measurements at large accelerators came to importance.

For the purposes of this study, particle accelerators can be considered as sequences of linear focusing elements (magnetic lenses) arranged either in a circle (circular accelerators) or in a line (linear accelerators). In an ideal accelerator with perfectly aligned magnetic elements, the orbit of the charged particle beam passes through the centers of the magnetic lenses. Any alignment error results in beam orbit distortion. If the distortions are large compared to the apertures of the lenses or the size of the vacuum chambers or the size of linear focusing field areas, they become an obstacle to the successful operation of the machine and must be corrected. This can be done either with the use of electromagnetic orbit correctors or by means of mechanical realignment which brings the centers of the focusing lenses back to their ideal positions. In large accelerators with hundreds of magnetic elements, such as ones discussed in this Letter, the motion of the ground and the corresponding displacements of the magnets are the most important source of beam orbit distortions [5]. The larger effect is produced by the uncorrelated relative motion of the neighboring focusing elements while very long-wavelength movements are practically unimportant [6]. Typically, the ground motion effects start to be of a serious concern for accelerators at the amplitudes of the uncorrelated motion from a fraction of a micron to tens of microns [7]. For accelerators which produce collisions through the interaction of extremely small size beams, the final focusing magnet stability tolerances could be as tight as microns to a few nanometers [8]. Because of the concerns with the magnet position stability, large accelerators have usually been installed inside deep concrete-and-steel enforced tunnels. The typical diameters of the tunnels are in the range of 5-8 m at depths of 10 to 100 meters at sites with known good and stable geology.

Despite having sophisticated beam orbit correction systems, all accelerators undergo regular realignment of the magnets positions back to their ideal values. Such realignments allow to keep the obits within the range of the correction systems and helps to maintain stable operation of the facilities over periods of many years. Modern commercial instruments, e.g. laser trackers and digital levels, for geodetic survey and alignment allow one to achieve accuracies of a fraction of a mm over distances of a km. Their description can be found in [9, 10].

Hydrostatic level sensors (HLS) are routinely used at geophysics facilities [11,12], but usually in small numbers. High precision HLS probes have been developed and used in large numbers at the various accelerators at Fermi National Accelerator laboratory (FNAL, Chicago, US) and other locations in Illinois to study the diffusion in space or spatial correlations of the ground motion [13]. They employ capacitive sensors of the water levels and are equipped with local water temperature meters needed for thermal expansion compensation. The probes are made in two configurations – one for use with a single 1” diameter half-filled water pipe, and another for use with two separate ½” diameter tubes for air – to assure the same air pressure inside all the probes - and for water (fully filled). A pair of the probes set side-by-side shows a differential noise level with rms value increasing with the time interval $T$ as $\sigma^2=(0.09\mu m)^2+ 1.252\cdot 10^{-7}\ \mu m^2/s\cdot T[s]$. In a typical measurement arrangement, six to twenty such probes spaced 15 to 30 meters apart are installed in the same water level system. Once a minute a PC based data acquisition system collects not only the water level data (averaged over a minute), but also all of the probe's temperature readings for the data correction and readings from one or two air pressure monitors.

The beam position in typical accelerators is monitored at many locations with very good accuracy, e.g. in the Tevatron Collider at FNAL – by 240 electrostatic pickup electrodes with intrinsic resolution of about 5 μm [14]. The orbit data are acquired either routinely at a 1 Hz sampling rate or by request at faster rates, stored and made available for processing.

As mentioned above, the diffusive motion of the ground is often just a background to much more powerful processes, like ground expansion due to temperature changes, or bending due to atmospheric pressure variation or winds, long-term settlement drifts or Earth tides. Special data processing is often needed to separate diffusive noise from systematic or periodic signals. In the time- or space-domains, that can be achieved with the use of digital filters, e.g. for a raw signal $dY(t)$ one can compute and analyze either the first difference $dY(t)-dY(t+T)$ or the second difference $dY(t)+dY(t+2T)-2dY(t+T)$ each of which is effectively a high frequency filter that cuts out linear trends and slow periodic variations leaving the noise component intact. In the frequency- or wavelength- domains, the power spectra densities of ground motion data usually contain peaks due to the periodic components which can be easily separated from a power-law component of the spectrum due to random processes. Comprehensive description of the geophysics time series analysis and methods can be found in [1].

The Tevatron Collider in Batavia, IL (USA) is one of the world's highest energy accelerators for elementary particles research with beams of 980 GeV protons and antiprotons circulating in opposite directions inside the set of 774 bending magnets and 216 focusing magnets regularly spaced in the 6.3 km long tunnel ring at approximately 7 m below the surface. The motion of tunnel floor translates into motion of the focusing magnets and that translates further into movement of the beams. For effective operation of the Collider, the beam orbit motion must be stabilized to within 0.1mm by means of an automatic orbit correction system. Without such a system daily changes in the orbit position could easily reach 0.2-0.3 mm and as much as 0.5-1 mm over the periods of 2-4 weeks [14]. The alignment system of the Tevatron employs more than 200 geodetic "tie rods" (thick metal rods screwed into the concrete tunnel wall all over the ring and equipped to hold spherical retroreflectors for precise position measurements), each spaced approximately 30 m apart.

The positions of the magnets are regularly referenced locally with respect to the "tie rods" while the positions of all the "tie rods" are routinely monitored. The "tie rod" elevation data sets are available for the years of 2001, 2003, 2005, 2006 and 2007. Fig.1 shows the change of the elevations around the ring accumulated over two intervals – 2 years (2003-2005) and 6 years (2001-2007). One can see that longer term motion has a larger amplitude. The variance $<dY^2(L)>=<(dY(z)-dY(z+L))^2>$ of the elevation difference of the points as a function of the lag (distance between pairs of the measurement points) $L$ has been calculated and averaged over all possible time intervals. That is to say, there are two 1-year intervals (2005-2006, 2006-2007), three 2-year intervals (2001-2003, 2003-2005, 2005-2007), etc, and one for the 6-year interval 2001-2007. The results for the 1-year changes and for the 6-year change are shown in Fig.2. A remarkable difference between the two plots is that 1 year variance scales linearly only up to $L\approx$700-800 m and does not depend on $L$ beyond that scale, while the 6 years variance grows all the way to distances as large as 1800 m. The linear dependence on $L$ is indicative of a significant level of interdependence of the movements of distant points. The calculated variances for all possible time differences can be well approximated by linear fits $<dY^2(L)> =a+bL$ over distances less than 900 m and the slopes (fit parameters $b$ with the error bars) are plotted in Fig.3.

One can see that the variance per unit distance grows with the time interval between the measurements, and can be approximated by a linear fit $b(T) = cT$ with $c=0.153\pm0.004$ $[mm^2/km/year]$. The Tevatron “tie rod” data presented in Figs.2-3can be consolidated in an empirical $ATL$ law [15]:

$$<dY^2>=ATL$$

with coefficient $A_{Tevatron} = c = (4.9\pm0.13)\cdot10^{-6}$ μm$^2$/s/m. This relation is characteristic of a “random-walk” process, or diffusion, both in time and in space. Note, that for independent movements of the ground elements $<dY^2>=const$ .

Similar diffusion rates have been measured by a system of 20 hydrostatic level sensor (HLS) probes installed on top of the Tevatron focusing magnets and spaced 30 m apart and connected by a half-filled water pipe. For this system, the data for the lags $L\leq120$ m and $T\leq1$ week can be approximated by an $ATL$ law with coefficient $A_{TevHLS} = (2.2\pm1.2)\cdot10^{-6}$ μm$^2$/s/m [16]. For the lags larger than 120 m, the variance of the displacements accumulated over 1 week does not depend on the lag. The characteristic time dependence of the $ATL$-like diffusion has been observed in beam orbit drifts in many particle accelerators [16]. For example, the Tevatron beam orbits are being measured with micron precision in hundreds of locations around the ring and found to wander from ideal positions due to motion of focusing magnets ( distortions from numerous uncorrelated magnet moves add in quadrature). In addition to the 12- and 24-hour variations associated with the tides and daily temperature effects, the orbit motion has a diffusive component which corresponds to coefficients $A_{Tevatron\ V} = (2.6 \pm 0.3) \cdot 10^{-6}$ μm$^2$/s/m and $A_{Tevatron\ H} = (1.8 \pm 0.2) \cdot 10^{-6}$ μm$^2$/s/m (different in vertical and horizontal planes).

Data from more than two dozen measurements made at the Tevatron and several other large accelerators as well as results of similar studies made elsewhere has been analyzed in [16] and are summarized in Table I. These measurements employed a variety of instruments: beam position monitors to observe orbit drifts in accelerators, modern laser trackers and digital levels to do geodetic surveys of magnets, laser interferometers and HLS systems are used in geophysics studies. The calculated diffusion coefficients *A* are given in the third column in Table I, the second column indicates whether the diffusion has been observed in the time domain (*T*) or in the space domain (*L*) or simultaneously in both (*T*,*L*). More details (e.g., the depth of the tunnel/measurement site, spacing $\Delta L$ between measurement points, etc) as well as all corresponding references can be found in [16]. The diffusion rates measured at the same site by different methods are in reasonable agreement with each other. The diffusion coefficients have a tendency to be smaller at greater depths, in harder rocks and in geologically stable locations. There are indications that the methods of tunneling – boring vs. blasts - may affect the diffusion rate.

The measurements presented above unambiguously show that ground motion is not a random stochastic uncorrelated noise. The observed "space-time random walk" nature is an indication of fractal dynamics of cascades of geological blocks of various sizes (a possible model is suggested in [16]).

Naturally, for small time intervals the movements of the ground elements is fully uncorrelated if they are separated by long enough distance – for example, by more than 120 m for 1 week intervals as seen in the Tevatron HLS data or by more than 800 m for 1 year intervals as seen in the Tevatron alignment data discussed above. More detail exploration of such a boundary between the *ATL*-like and the fully uncorrelated regimes will provide further insight into the dynamics of the ground fractures.

In summary, high precision measurements of the movements of the large accelerator tunnels over the range of spatial scales from several meters to tens of km and time intervals from minutes to several years show that the diffusive motion of ground elements has a characteristic "random-walk" nature both in time and in space, i.e., looks like a convolution of two Brownian processes - one in the space-domain and another in the time-domain. That indicates the fractal dynamics of cascades of geological blocks of various sizes. The data can be approximated by a simple empirical formula $<dY^2>=ATL$ which allows to estimate the long-term movements of accelerator tunnels and other large scale constructions as long as the site dependent diffusion constant $A$ is determined.

The author is very thankful to many people from accelerators worldwide who provided me with raw data record for further ground diffusion analysis [16]. My special thanks to V.Parkhomchuk who brought my attention to the deep physics issues associated with ground motion and was the first who coined the term "*ATL*-law". Fermi National Accelerator Laboratory is operated by Fermi Research Alliance, LLC under Contract No. DE-AC02-07CH11359 with the United States Department of Energy.

[1] B.D.Malamud and D.L.Turcotte, Advances in Geophysics **40**, 1 (1999)

[2] *Fractals in the Earth Sciences*, edited by C.C.Barton and P.R.LaPointe (Springer, 1995)

[3] D.L.Turcotte, *Fractals and Chaos in Geology and Geophysics* (Cambridge University Press, 1997)

[4] D.C.Agnew, Geophysical Research Letters **19**, 333 (1992).

[5] G.E.Fischer, in *Summer School on High Energy Particle Accelerators*, Batavia (1984), AIP Conf. Proc. No.153 (AIP, New York, 1987), pp. 1047-1119

[6] V.Parkhomchuk, V.Shiltsev and G.Stupakov, Particle Accelerators, **46**, 241 (1994).

[7] B.Baklakov, *et al.,* Phys. Rev. ST Accel. Beams **1**, 031001 (1998)

[8] A.Seriy and O.Napoly, Phys. Rev. E **53**, 5323 (1996)

[9] see e.g. series of *Proceedings of Int. Workshop on Accelerator Alignment (IWAA)* at http://www-conf.slac.stanford.edu/iwaa/

[10] J.Berger, in *Advances in Geophysics* **16,** edited by H.E.Landsberg and J. van Mieghem (Academic Press, Inc., New York, USA, 1973)

[11] D.C.Agnew, Rev. Geophys. **24**, 579 (1986)

[12] N.D'Oreye and W.Zuern, Rev. Sci. Instrum. **76**, 024501 (2005)

[13] A.Seryi, *et al.*, in *Proceedings of 2001 IEEE Particle Accelerator Conference (PAC 2001),* Chicago, USA, (IEEE, Piscataway, NJ, 2001), p.1479.

[14] R.Moore , A. Jansson and V.Shiltsev , Journal of Instrumentation JINST **4** , P12018 (2009)

[15] B.Baklakov, *et al.*, Technical Physics, **38(10)** , 894 (1993); translated from Sov. Zh. Tech. Fiz., **63**, No.10, 123-132 (1993).

[16] V.Shiltsev, Preprint FERMILAB-FN-0834-APC (FNAL, 2009), arXiv:0905.4194

TABLE I. Ground diffusion coefficients *A* measured at different sites (from [16], (V) – vertical, (H) – horizontal plane). The fourth column indicates the maximum length of time record, while the fifth is either the circumference of the circular accelerators or maximum length of the measurement system.

| | | *A*, $10^{-6}$ $\mu m^2/s/m$ | Time | Scale |
|---|---|---|---|---|
| *Tevatron Collider Data* | | | | |
| "Tie-rods" (V) | *L,T* | 4.9±0.1 | 1-6 yr | 6.3km |
| 20 HLS system | *L,T* | 2.2±1.2 | 1 week | 600m |
| Beam Orbit (V) | *T* | 2.6±0.3 | 15 hrs | 6.3km |
| (H) | *T* | 1.8±0.2 | 15 hrs | 6.3km |
| *Beam Orbit Drifts in Other Accelerators* | | | | |
| HERA -*e* (V) | *T* | 4±2 | 25 days | 6.3km |
| HERA -*p* (V) | *T* | 8±4 | 5 days | 6.3km |
| TRISTAN (V) | *T* | 27±7 | 2 days | 3.0km |
| Circmf. KEKB | *T* | 27±3 | 4 mos. | 3.0km |
| LEP (V) | *T* | 10.9±6.8 | 18hrs | 26.7km |
| LEP (V) | *T* | 39±23 | 3.3hrs | 26.7km |
| (H) | *T* | 32±19 | 3.3hrs | 26.7km |
| SPS (V) | *T* | 6.3±3.0 | 2 hr | 6.9km |
| *Accelerator Alignment/Survey Data Analysis* | | | | |
| CERN LEP (V) | *L,T* | 6.8-9.0 | 6,9 mos | 26.7 km |
| | | 3±0.6 | 6 years | 26.7km |
| CERN SPS (V) | *L,T* | 14±5 | 3-12 yr | 6.9km |
| *Ground Motion Studies Data* | | | | |
| PFO (CA, USA) | *T* | 0.7 | 5 years | 732 m |
| SLAC Linac (V) | *T* | 1.4±0.2 | 0.5 hr | 3 km |
| | *T* | 0.2-2 | 1 hr | 3 km |
| Esashi (Japan) | *T* | 0.3-0.5 | 15 years | 50 m |
| Sazare (Japan) | *T* | 0.01-0.12 | 6 weeks | 48 m |
| KEKB tunnel | *T* | 40 | 4 days | 42 m |
| FNAL PW7 | *T* | 6.4±3.6 | 3 months | 180 m |
| FNAL MINOS | *T,L* | 0.18 | 1 month | 90m |
| Aurora mine | *T,L* | 0.6±0.3 | 2 weeks | 210m |

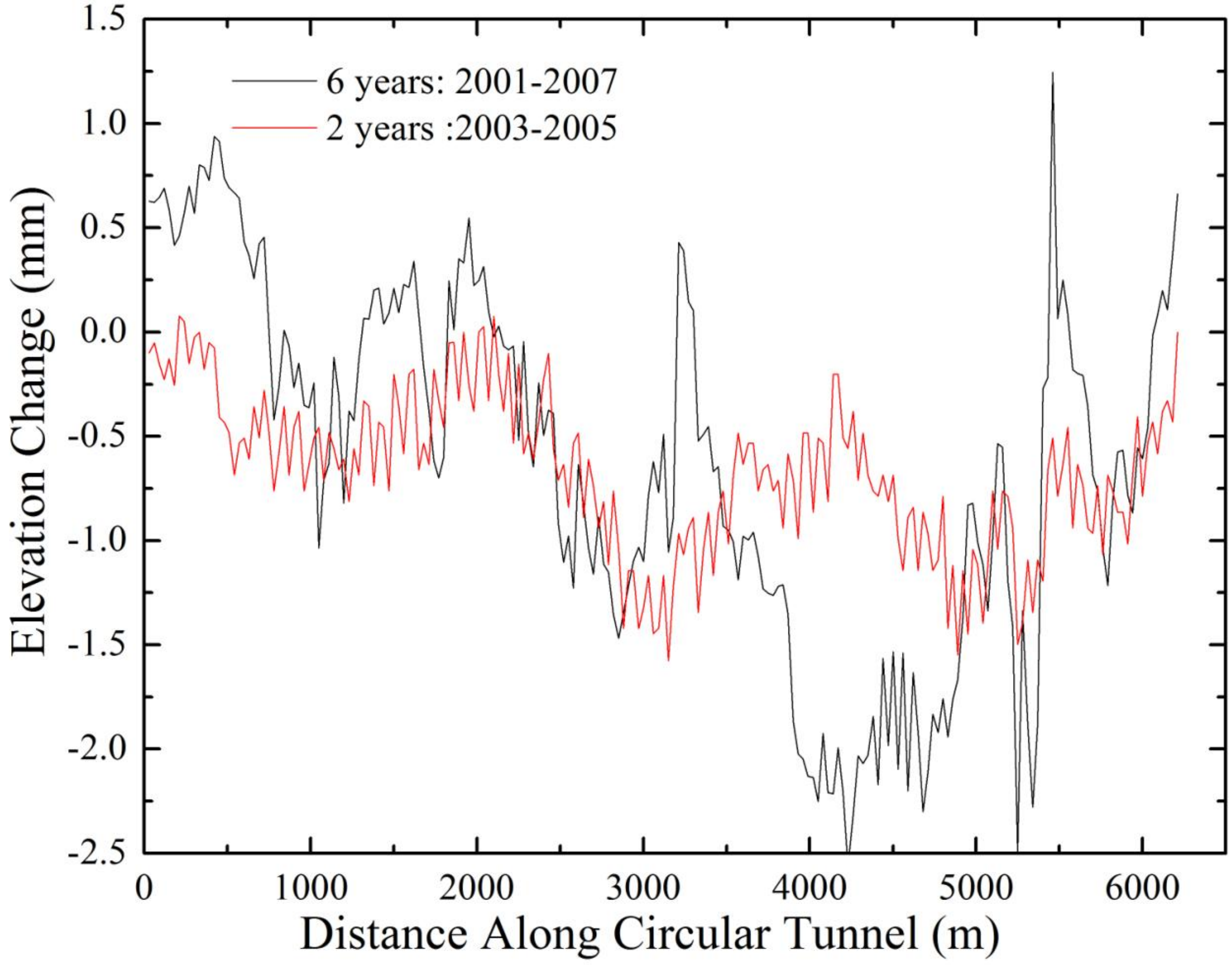

FIG.1: Vertical displacement of more than 200 “tie rods” in the Tevatron tunnel over the period 2003-2005 and a 6 year period of 2001-2007.

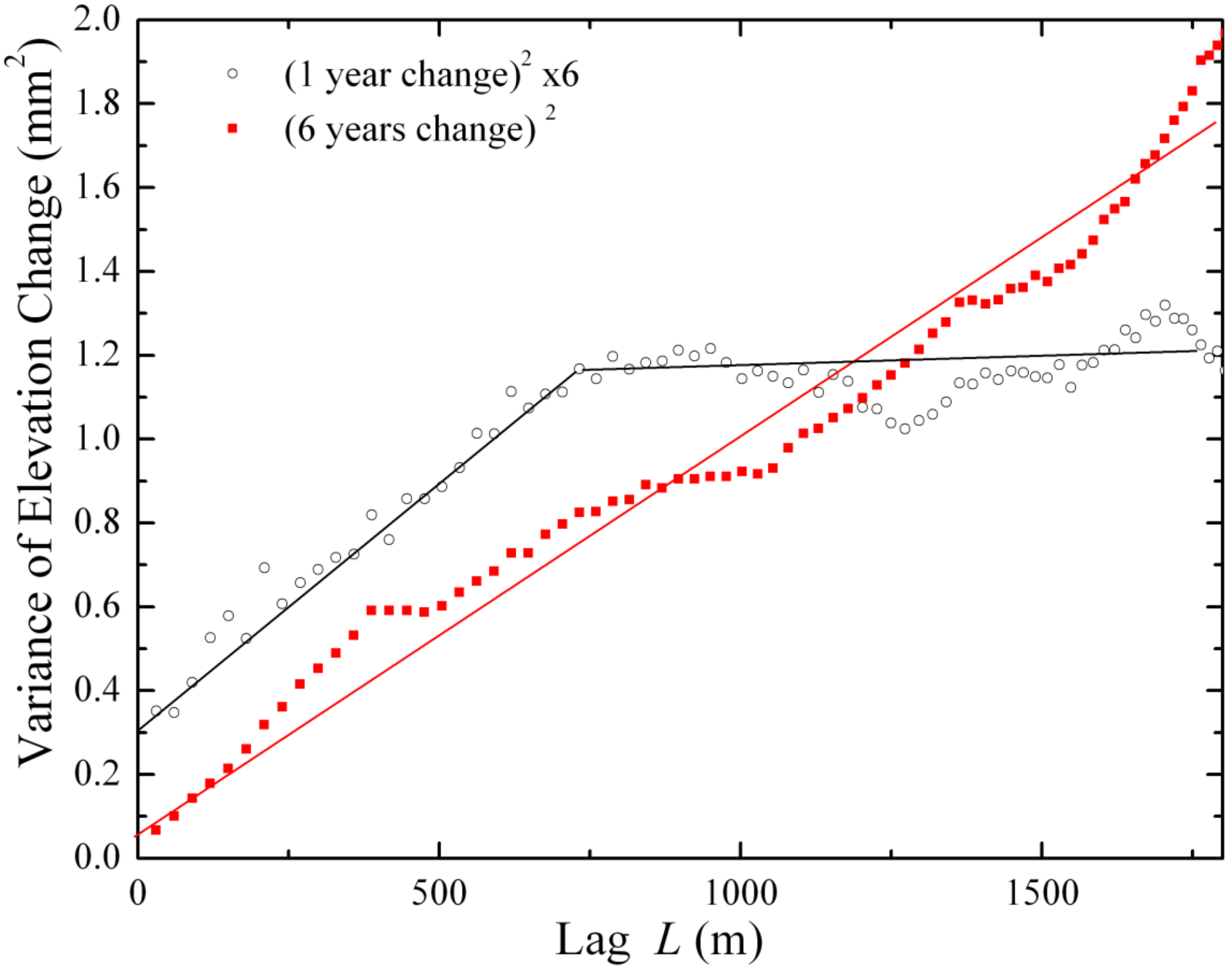

FIG.2: Variances of the Tevatron "tie rod" vertical displacements over time intervals of 1 year (multiplied by 6) and 6 years vs the distance *L*.

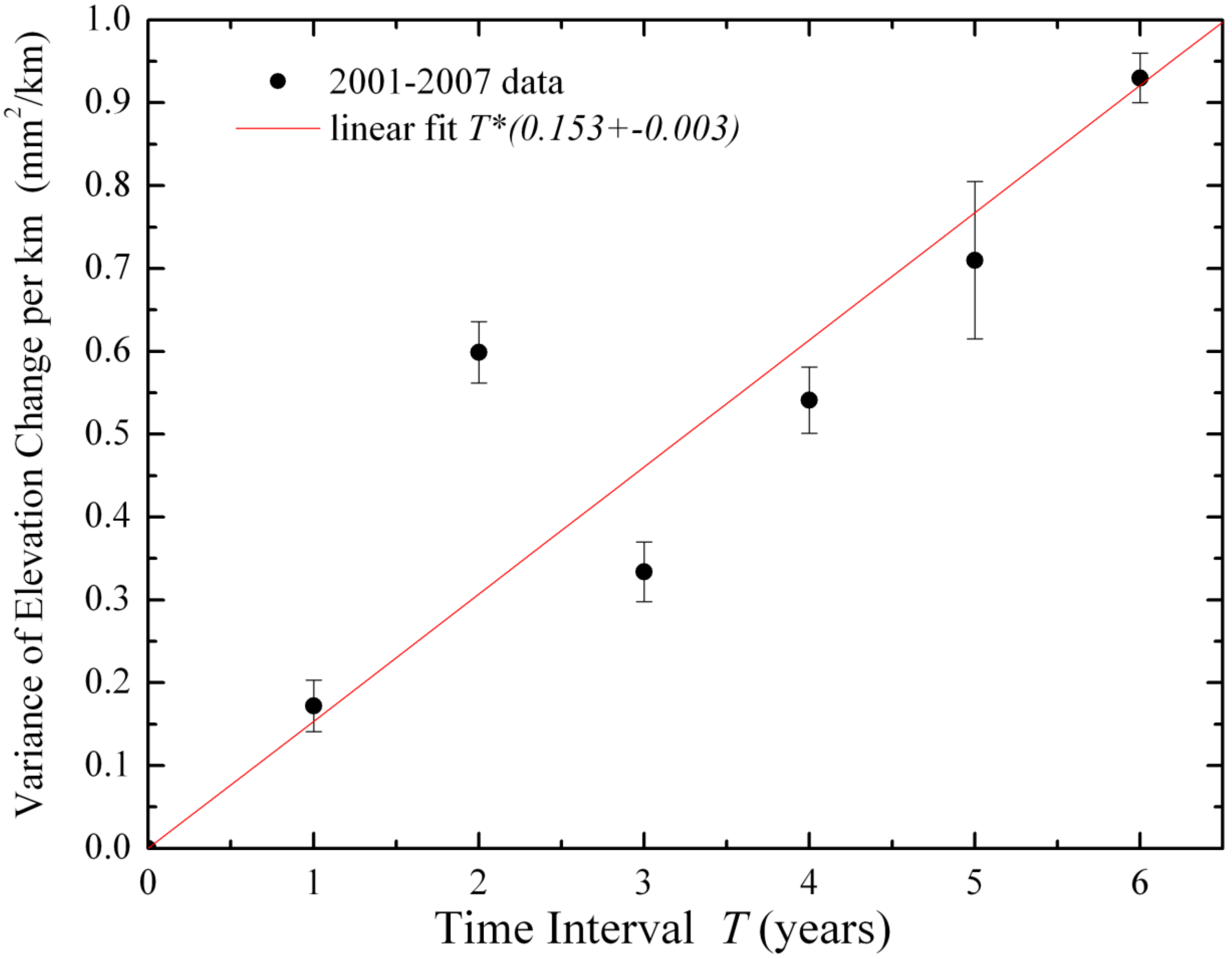

FIG.3: Variances of the Tevatron alignment "tie rods" displacements per unit distance vs the time interval between the measurements.